\documentclass[aps,superscriptaddress,prc,twocolumn,nofootinbib]{revtex4-1}

\usepackage{graphicx}
\usepackage{epstopdf}
\usepackage{subfigure}
\usepackage{epsfig}
\usepackage{amsmath,amssymb,amsfonts}
\usepackage[utf8]{inputenc}
\usepackage{color}
\usepackage[bookmarks,
                bookmarksopen = true,
                bookmarksnumbered = true,
                linktocpage,
                colorlinks = true,
                linkcolor = blue,
                urlcolor  = blue,
                citecolor = blue,
                anchorcolor = green,
                hyperindex = true,
                hyperfigures]
                {hyperref}

\begin{document}

\title{Nuclear modification of leading hadrons and jets within a virtuality ordered parton shower}

\author{Shanshan Cao}
\affiliation{Department of Physics and Astronomy, Wayne State University, 666 W. Hancock St., Detroit, MI 48201, USA}
\author{Abhijit Majumder}
\affiliation{Department of Physics and Astronomy, Wayne State University, 666 W. Hancock St., Detroit, MI 48201, USA}

\date{\today}


\begin{abstract}

The event generator based on the higher-twist energy loss formalism -- {\it Modular All Twist Transverse-scattering Elastic-drag and Radiation} (\textsc{Matter}) -- is further developed and coupled to a hydrodynamic model for studying jet modification in relativistic nuclear collisions. The probability of parton splitting is calculated using the Sudakov form factor that is constructed by a combination of vacuum and medium-induced splitting functions; and the full parton showers are simulated, including both energy-momentum and space-time evolutions of all jet partons. With the assumption that partons below a virtual scale of 1~GeV is absorbed by the medium, this framework is able to provide a reasonable description of the nuclear modification of both leading hadrons and jets at high transverse momentum at RHIC and the LHC. 

\end{abstract}

\maketitle


\section{Introduction}
\label{sec:Introduction}

The nuclear modification of energetic hadrons and jets serves as a crucial signature of the formation of the color deconfined quark-gluon plasma (QGP) matter in heavy-ion collisions at the Relativistic Heavy-Ion Collider (RHIC) and the Large Hadron Collider (LHC)~\cite{Wang:1991xy, Qin:2015srf}. The suppression of single inclusive hadron and jet spectra at large transverse momentum ($p_\mathrm{T}$) is commonly understood as the consequence of in-medium scatterings experienced by high energy partons produced via initial hard scatterings as they travel through and interact with the QGP before fragmenting into hadrons~\cite{Braaten:1991we, Wang:1991xy, Gyulassy:1993hr, Baier:1996kr, Zakharov:1996fv, Qin:2015srf}. Phenomenological studies of parton energy loss involve sophisticated calculations of the medium modification of single inclusive hadron~\cite{Bass:2008rv, Armesto:2009zi, Chen:2010te,Cao:2017hhk,Arnold:2001ba,Wang:2001ifa,Salgado:2003gb,Marquet:2009eq,Dainese:2004te,Vitev:2002pf,Vitev:2004bh,Armesto:2005iq,Renk:2011gj,Horowitz:2011gd,Chen:2011vt,Majumder:2010ik,Renk:2010mf,Wicks:2005gt}, di-hadron~\cite{Majumder:2004pt,Zhang:2007ja,Renk:2008xq,Cao:2015cba}, $\gamma$-hadron correlation~\cite{Zhang:2009rn,Qin:2009bk,Wang:2013cia,Chen:2017zte}, full jets~\cite{Qin:2010mn,Dai:2012am,Wang:2013cia,Chang:2016gjp,Casalderrey-Solana:2014bpa}, as well as groomed jet substructures~\cite{Chien:2016led,Mehtar-Tani:2016aco,Chang:2017gkt}. Jet-medium scatterings are governed by a series of jet transport coefficients, among which the most commonly quoted parameter is the jet transport coefficient $\hat{q}$ that denotes the transverse momentum transfer squared per unit length $d\langle p_\perp^2\rangle/dt$ between the propagating hard parton and the soft medium. This jet transport coefficient is a local quantity and could depend on local temperature, jet energy and virtuality scales, as well as both the coupling of the jet partons with the medium and the interactions between constituents within the medium itself. A systematical extraction of $\hat{q}$ was performed by the JET Collaboration by comparing several jet quenching model calculations to the experimental data of the nuclear modification factor ($R_\mathrm{AA}$) of high $p_\mathrm{T}$ hadrons at RHIC and the LHC \cite{Burke:2013yra}. 

Monte-Carlo event generators are essential for quantitative theoretical studies of jet observables. They naturally include multi-particle effects of the full parton showers, and take into account the fluctuations of parton energy loss as well as the QGP medium. Thus, they represent an important interface between theory and experiment. While event generators are very well developed for parton showers inside vacuum~\cite{Sjostrand:1986hx,Sjostrand:2006za,Marchesini:1991ch,Gleisberg:2008ta}, consensus has not been reached for jet modification with the presence of a medium. This is mostly due to the lack of a unified theoretical approach that covers the full phase space of jet evolution in heavy-ion collisions. Various formalisms of parton energy loss have been developed based on different assumptions of the phase space~\cite{Guo:2000nz,Gyulassy:2000fs,Wiedemann:2000za,Arnold:2002ja}, and one may refer to Ref.~\cite{Armesto:2011ht} for a detailed comparison between different approaches. Transport models, such as \textsc{Lbt}~\cite{Cao:2016gvr,Cao:2017hhk,Chen:2017zte} based on the higher-twist scheme~\cite{Guo:2000nz} and \textsc{Martini}~\cite{Schenke:2009gb} based on the the Arnold-Moore-Yaffe (AMY) scheme~\cite{Arnold:2002ja}, have been developed to describe parton splittings at low virtuality scale. These transport models usually start with vacuum parton showers generated by \textsc{Pythia}~\cite{Sjostrand:2006za} and simulate their subsequent evolution through the medium. An alternative way of combining medium modification with vacuum parton showers is to introduce medium-induced scatterings of jet partons between two vacuum splittings inside \textsc{Pythia}, as implemented by \textsc{Jewel}~\cite{Zapp:2008gi,Zapp:2012ak}. 

Without simulating parton showers using an event generator, one may also calculate hadron spectra in nucleus-nucleus collisions by convoluting the medium-induced gluon spectra, parton fragmentation function with the initial parton spectra in proton-proton collisions, as implemented in the CUJET model~\cite{Xu:2014ica,Xu:2015bbz} that is based on the Djordjevic-Gyulassy-Levai-Vitev (DGLV) energy loss formalism~\cite{Gyulassy:2000er,Djordjevic:2004nq}, and other semi-analytical calculations based on AMY~\cite{Qin:2007rn}, higher-twist~\cite{Majumder:2011uk} and Armesto-Salgado-Wiedemann (ASW)~\cite{Salgado:2003gb} formalisms. In all cases above, the splittings of partons at high virtuality are vacuum-like and not modified by the medium. On the contrary, the medium modification of parton showers at high virtuality is typically described using medium-modified DGLAP evolution~\cite{Kang:2014xsa,Chien:2015vja,Majumder:2011uk}. Such DGLAP-type of parton showers within a medium can be simulated via event generators such as \textsc{Q-Pythia}~\cite{Armesto:2009fj} that directly modifies the Sudakov form factor embedded in \textsc{Pythia} such that both vacuum and medium-induced splitting functions contribute to the splitting of virtual partons. In \textsc{Q-Pythia}, the medium-induced part is taken from the ASW~\cite{Armesto:2003jh} scheme.

Following our earlier study~\cite{Majumder:2013re,Kordell:2017hmi}, we develop a Sudakov-type parton shower model based the higher-twist energy loss formalism -- the {\it Modular All Twist Transverse-scattering Elastic-drag and Radiation} event generator (\textsc{Matter}) -- to simulate the splitting of highly virtual partons, i.e. partons whose virtuality (with a unit of GeV$^2$) $t  \gg \sqrt{\hat{q}E}$ where $E$ is the energy of the parton. At such high virtuality, the dominant mechanism of splitting is described using a medium-modified virtuality-ordered shower~\cite{Majumder:2009ge,Majumder:2009zu,Wang:2001ifa,Majumder:2011uk}, in which scatterings in the medium produce a small variation in the vacuum splitting function. The setup of the formalism ensures that the number of splittings dominates over the number of scatterings. Within this approach, vacuum and medium induced radiation is accounted for simultaneously, and the space-time structure (including its fluctuations) of the shower is introduced. This \textsc{Matter} event generator is coupled to a hydrodynamic medium for studying parton showers in the QGP, and provides reasonable descriptions of the nuclear modification of both high $p_\mathrm{T}$ hadrons and jets.

This paper is organized as follows. In Sec.~\ref{sec:matter}, we present an overview of constructing a virtuality-ordered parton shower model based on the Sudakov form factor, and validate our \textsc{Matter} event generator by comparing our parton spectra in vacuum showers with the well established \textsc{Jetset}~\cite{Sjostrand:1986hx} results. In Sec.~\ref{sec:results}, we use \textsc{Matter} to calculate the nuclear modification factor $R_\mathrm{AA}$ and elliptic flow coefficient $v_2$ of both single hadrons and jets, and compare them to experimental data at RHIC and the LHC. In Sec.~\ref{sec:summary}, we summarize and discuss an outlook of future development.

\section{The \textsc{Matter} event generator for parton showers}
\label{sec:matter}

For a hard parton produced at a point $r$ with a forward light-cone momentum $p^{+} = (p^0 + \hat{n}\cdot \vec{p})/\sqrt{2}$ ($\hat{n}=\vec{p}/|\vec{p}|$ represents the direction of the parton), one may construct the following Sudakov form factor that gives the probability of no splitting through a given channel $i$ ($q\rightarrow qg$, $g\rightarrow gg$ or $g\rightarrow q\bar{q}$) between virtuality scales $t$ ($t = Q^{2}$) and $t_\mathrm{max}$~\cite{Hoche:2014rga}:
\begin{equation}
\label{eq:sudakov}
\Delta_i (t_\mathrm{max},t)=\exp\left[- \int\limits_t^{t_\mathrm{max}}\frac{d\tilde{t}}{\tilde{t}}\frac{\alpha_\mathrm{s}(\tilde{t})}{2\pi}\int\limits_{z_\mathrm{c}}^{1-z_\mathrm{c}} dy P_i (y,\tilde{t})\right].
\end{equation}
Here, $u$, $d$ and $s$ are considered for quark flavors, and $z_\mathrm{c}=t_\mathrm{min}/\tilde{t}$ is taken from the kinematic constraints of splitting, where the minimum allowed virtuality is set as $t_\mathrm{min}=1~\mathrm{GeV}^2$. The splitting function $P_i$ is obtained by adding its vacuum and medium-induced contributions:
\begin{equation}
\label{eq:totP}
P_i(y,\tilde{t})=P_i^\mathrm{vac}(y)+P_i^\mathrm{med}(y,\tilde{t}),
\end{equation}
where the medium-induced part is adopted from the higher-twist energy loss calculations~\cite{Guo:2000nz,Majumder:2009ge,Aurenche:2008hm,Aurenche:2008mq} and treated as a perturbation on top of the vacuum part:
\begin{align}
\label{eq:medP}
P_i^\mathrm{med}&(y,\tilde{t})=\frac{P_i^\mathrm{vac}(y)}{y(1-y)\tilde{t}}\int\limits_0^{\tau_f^+}d\zeta^+ \hat{q}(r+\zeta)\Biggl[2-2\cos \left(\frac{\zeta^+}{\tau_f^+}\right) \nonumber \\
&-2\frac{\zeta^+}{\tau_f^+}\sin\left(\frac{\zeta^+}{\tau_f^+}\right)+2\left(\frac{\zeta^+}{\tau_f^+}\right)^2\cos \left(\frac{\zeta^+}{\tau_f^+}\right)\Biggl].
\end{align}
In Eq.~(\ref{eq:medP}), $\hat{q}$ is the gluon jet transport coefficient that denotes its transverse momentum broadening squared per unit length due to elastic scattering and is evaluated at the location of scattering $\vec{r}+\hat{n}\zeta^+$; $\tau^+_f=2p^+/\tilde{t}$ is the mean formation time of the splitting. In the case of the vacuum portion of the split, there is no scattering, and the splitting time and location have been integrated out to obtain a splitting kernel entirely in momentum space. In order for both the vacuum and medium modified kernels to be consistent, the location of scattering $\zeta^+$ is integrated only up to the mean formation time $\tau^+_f$.

Compared to the work of Guo and Wang (GW)~\cite{Guo:2000nz}, that of Aurenche-Zakharov-Zaraket (AZZ)~\cite{Aurenche:2008hm,Aurenche:2008mq} improves the calculation by taking into account of the transverse momentum from the medium within the phase factors in the higher-twist matrix elements. However, since only the emitted gluon scattering with the medium is considered, Eq.~(\ref{eq:medP}) is not positive definite when $\zeta^+$ is much larger than $\tau_f^+$. A positive definite splitting function may be expected after one includes the complete set of terms at next-to-leading twist. This will be discussed in details in our upcoming work. 

Regardless of the differences between GW and AZZ, as long as one integrates the medium modified splitting function up to a given length in the calculation of the Sudakov form factor, the different calculations only differ by an overall normalization factor. In the end, this only affects the $\hat{q}$ value one uses in model-to-data comparison. In this work, both values of $\hat{q}$ will be reported for applying the AZZ splitting function [Eq.~(\ref{eq:medP})] and the GW splitting function [only the $2-2\cos(\zeta^+/\tau_f^+)$ contribution in Eq.~(\ref{eq:medP})]. On integrating Eq.~\eqref{eq:medP} above up to $\zeta^+ = \tau^+_f$, we obtain that the GW result is 50\% higher than the AZZ result.
Note that Eq.~(\ref{eq:medP}) is also similar to the results given in Refs.~\cite{CasalderreySolana:2007sw, Ovanesyan:2011kn}, but not exactly the same because we absorb the medium information into the $\hat{q}$ parameter here instead of a specific assumption of the scattering centers for the medium.

The probability of no splitting through any channel is then obtained by
\begin{equation}
\label{eq:totsudakov}
\Delta(t_\mathrm{max},t)=\prod\limits_i \Delta_i(t_\mathrm{max},t).
\end{equation}
Thus, splitting of a given parton is allowed if $r>\Delta(t_\mathrm{max},t_\mathrm{min})$ is satisfied, where $r$ is randomly sampled within $(0,1)$. And the corresponding virtuality scale $t$ at which the parton splits can be obtained by solving $r=\Delta(t_\mathrm{max},t)$. The specific channel $i$ through which the parton splits is then determined by their branching ratios:
\begin{equation}
\label{eq:BR}
\mathrm{BR}_i(t)=\int\limits_{t_\mathrm{min}/t}^{1-t_\mathrm{min}/t}dy P_i(y,t).
\end{equation}
With a selected splitting channel, the momentum fraction $y$ shared by the two daughter partons is determined by the splitting function $P_i(y)$. On the other hand, if $r>\Delta(t_\mathrm{max},t_\mathrm{min})$ is not satisfied, we treat the parton stable and set $t=t_\mathrm{min}$.

To start with, the initial maximum possible virtuality $t_\mathrm{max}$ of a hard parton is set as its energy produced in the initial hard scattering. After the first splitting, $y^2 t$ and $(1-y)^2 t$ are used as the new upper limits $t_1^\mathrm{max}$ and $t_2^\mathrm{max}$ for the two daughter partons, from which their actual virtualities $t_1$ and $t_2$ are sampled. This completes the full splitting process
\begin{align}
\label{eq:splitting}
\biggl[p^+&,\frac{t}{2p^+},0\biggl]\rightarrow\\
\biggl[&yp^+,\frac{t_1+\vec{k}_\perp^2}{2yp^+},\vec{k}_\perp\biggl]+\biggl[(1-y)p^+,\frac{t_2+\vec{k}_\perp^2}{2(1-y)p^+},-\vec{k}_\perp\biggl],\nonumber
\end{align}
from which the transverse momenta of the daughters with respect to the parent can be obtained:
\begin{equation}
\label{eq:kT}
k_\perp^2=y(1-y)t-(1-y)t_1-yt_2.
\end{equation}
The location of the splitting (or where the two daughter partons are produced) is calculated via $\vec{r}+\hat{n}\xi^+$, where $\xi^+$ is sampled via a Gaussian distribution with a mean value of $\tau^+_f$:
\begin{equation}
\label{eq:tau}
\rho(\xi^+)=\frac{2}{\tau^+_f\pi}\exp\left[-\left(\frac{\xi^+}{\tau^+_f\sqrt{\pi}}\right)^2\right].
\end{equation}

Iteration of this splitting process generates a virtuality-ordered parton shower starting from a single parton with its initial $t_\mathrm{max}$ to a number of final-state partons with $t=t_\mathrm{min}$.

\begin{figure}[tb]
  \epsfig{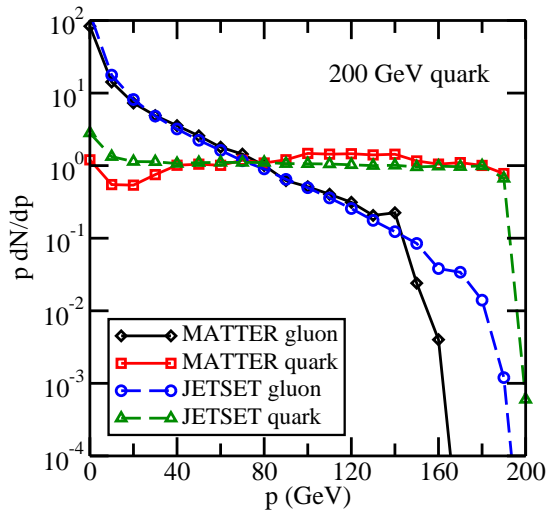}
  \caption{(Color online) Quark and gluon spectra from vacuum showers of a 200~GeV quark, compared between \textsc{Matter} and \textsc{Jetset}.}
 \label{fig:plot-JETSET}
\end{figure}

To validate our numerical setup, we compare the spectra of daughter quarks and gluons generated by \textsc{Matter} vacuum showers to those by \textsc{Jetset}~\cite{Sjostrand:1986hx} with angular ordering switched off in Fig.~\ref{fig:plot-JETSET}. The parton showers are initiated with single quarks with a fixed energy at 200~GeV and evolve in \textsc{Matter} with only vacuum contribution to the splitting function Eq.~(\ref{eq:totP}). As shown in Fig.~\ref{fig:plot-JETSET}, the parton spectra given by \textsc{Matter} simulation is consistent with those from \textsc{Jetset}. The \textsc{Matter} results were also compared to the semi-analytical solutions of the DGLAP equation in our earlier work~\cite{Majumder:2013re} for both vacuum and medium-modified parton showers, and shown consistent, although not exactly the same due to the somewhat different dynamics that is naturally included in a Monte-Carlo simulation versus a DGLAP evolution calculation.

To simulate parton showers in realistic heavy-ion collisions, we couple \textsc{Matter} to the dynamically expanding QGP matter simulated with a hydrodynamic model. In this work, we adopt the (2+1)-dimensional viscous hydrodynamic model \textsc{vishnew} developed in Refs. \cite{Song:2007fn,Song:2007ux,Qiu:2011hf}. The QGP fireballs are initialized with the Monte-Carlo Glauber model for their initial entropy density distribution. The starting time of the QGP evolution is set as $\tau_0=0.6$~fm and the specific shear viscosity ($\eta/s$=0.08) is tuned to describe the spectra of soft hadrons emitted from the QGP at both RHIC and the LHC. In this work, smooth averaged initial conditions are used for the hydrodynamical evolution. Possible effects of the initial state fluctuations on hard probes observables were discussed in our earlier work \cite{Cao:2014fna,Cao:2017umt} and shown small. The hydrodynamic simulation provides the spacetime evolution profiles of the local entropy density ($s$) and flow velocity ($u$) of the QGP. While the jet partons are inside the dense nuclear matter, i.e., the local temperature of the surrounding medium is greater than 160~MeV, both the vacuum and medium-induced parts contribute to the splitting function Eq.~(\ref{eq:totP}). We adopt a minimal assumption of the jet transport coefficient $\hat{q}$ in this work, where it is proportional to the local entropy density in the local rest frame of the fluid cell: $\hat{q}_\mathrm{local} = \hat{q}_0 \cdot s/s_0$, in which by convention $\hat{q}_0$ denotes the initial gluon transport coefficient in central $\sqrt{s_\mathrm{NN}}=200$~GeV Au-Au collisions at RHIC where $s_0$ is around 96~fm$^{-3}$. This $\hat{q}_0$ serves as the single parameter in the \textsc{Matter} calculation. The path length integral for calculating the medium-induced splitting function Eq.~(\ref{eq:medP}) is implemented in the center-of-mass frame of collisions. Effects of the local fluid velocity of the expanding medium are taken into account by utilizing the rescaled jet transport coefficient $\hat{q}=\hat{q}_\mathrm{local}\cdot p^\mu u_\mu/p^0$~\cite{Baier:2006pt} in Eq.~(\ref{eq:medP}). On the other hand, before jet partons enter the thermal medium ($\tau<0.6$~fm) or after they exit the dense nuclear matter, $\hat{q}$ is taken as 0 and thus only the vacuum splitting function contributes to parton showers. Within the \textsc{Matter} framework, elastic drag can be applied to parton showers as well by introducing the drag coefficient $\hat{e}$, but is beyond our discussion in this work.

\section{Nuclear modification of single hadron and jet production}
\label{sec:results}

In this section, we study the medium modification of single inclusive hadron and jet production in relativistic heavy-ion collisions.

Energetic partons are produced via hard scatterings at the early stage of heavy-ion collisions. Therefore, we initialize their production vertices with the Monte-Carlo Glauber model in which the nucleon-nucleon scattering cross sections are taken as 42~mb at $\sqrt{s_\mathrm{NN}}=200$~GeV and 64~mb at $\sqrt{s_\mathrm{NN}}=2.76$~TeV. The momentum distribution of energetic partons are calculated using the leading-order perturbative QCD (LO pQCD) calculation~\cite{Combridge:1978kx} convoluted with parton distribution functions via the CTEQ parametrizations~\cite{Lai:1999wy}. The LO pQCD cross sections are multiplied by a $K$-factor of 1.7 to take into account of additional contributions from higher-order corrections that increase the total yields of hard partons but barely affect the shapes of their spectra at high $p_\mathrm{T}$. The rapidity distributions of initial hard partons are assumed to be uniform in the mid-rapidity region ($-1<y<1$). Each energetic parton starts with a maximum possible virtuality $t_\mathrm{max}$ (we adopt $t_\mathrm{max} = E^2$ as a natural upper boundary for a time-like parton) and then evolve through \textsc{Matter} until all daughter partons approach $t_\mathrm{min} = 1~\mathrm{GeV}^2$. In proton-proton collisions, we only apply the vacuum splitting function to parton showers; while in nucleus-nucleus collisions, the vacuum plus medium-induced splitting function is utilized in which the jet transport coefficient $\hat{q}$ is calculated based on the local entropy density and flow velocity of the QGP given by the hydrodynamic simulation. Note that the Sudakov-based parton splitting is only valid for highly virtual partons. Thus, we restrict our application of \textsc{Matter} to partons with $t>t_\mathrm{min}$. If partons fall below $t_\mathrm{min}$ before they exit the QGP (more than 1~fm away from the medium boundary), they are regarded as part of the medium in this work. In principle, these less virtual partons may continue interacting with the QGP through on-shell transport models. This will be included in our upcoming study within the framework of multi-stage jet evolution~\cite{Cao:2017zih,Cao:2017crw}. At the end of \textsc{Matter} evolution, each final state parton is converted into hadron through \textsc{Pythia} simulation~\cite{Sjostrand:2006za} where the independent fragmentation mechanism is applied. This provides reasonable results for energetic single hadron production. On the other hand, since jets may include soft hadrons that are beyond the description of \textsc{Pythia} fragmentation, we construct jets using the anti-$k_\mathrm{T}$ algorithm at the parton level in this work. Jet reconstruction at the hadron level will be implemented in our future effort after the coalescence mechanism~\cite{Han:2016uhh} for soft hadron production is included.  

\begin{figure}[tb]
  \epsfig{file=plot-pp_ratio.eps, width=0.45\textwidth, clip=}
 \caption{(Color online) Single hadron and jet spectra in proton-proton collisions at RHIC~\cite{Adare:2007dg} and the LHC~\cite{CMS:2012aa,Khachatryan:2016jfl}: (a) for a direct comparison and (b) for the ratio between theory calculation and experimental data.}
 \label{fig:plot-pp_base}
\end{figure}

\begin{figure}[tb]
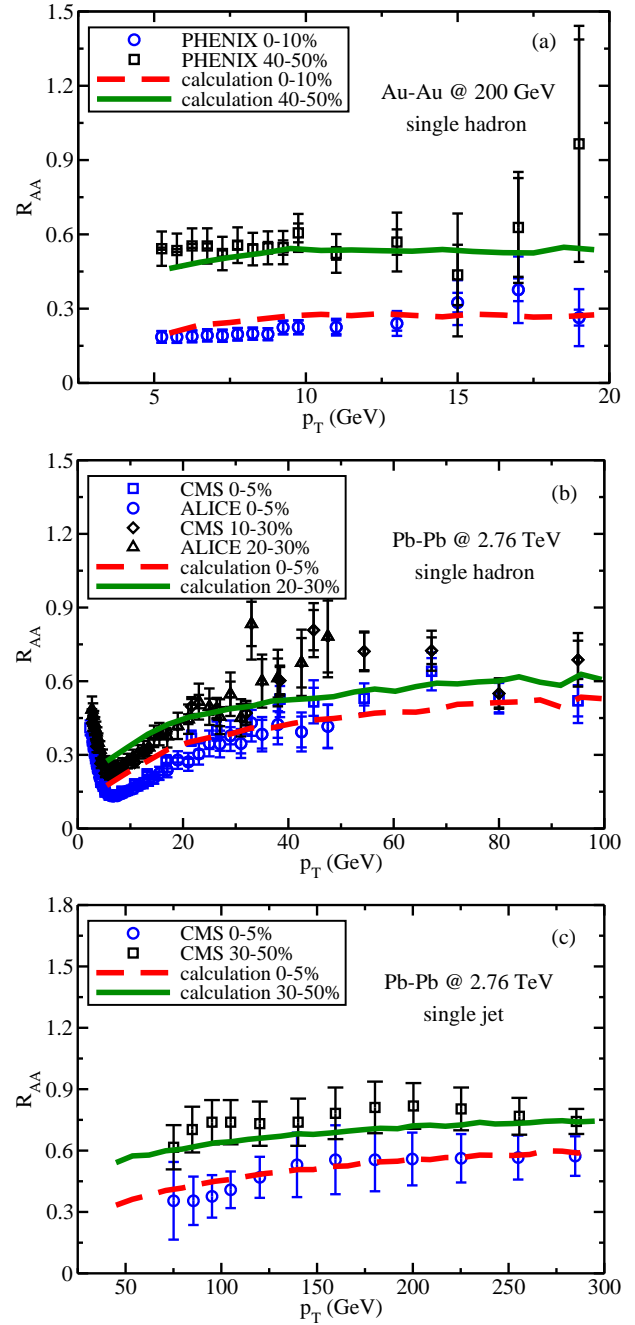

 \subfigure{\label{fig:plot-RAA-pion-200}
  \epsfig{file=plot-RAA-pion-200.eps, width=0.45\textwidth, clip=}}
 \subfigure{\label{fig:plot-RAA-pion-2760}
  \epsfig{file=plot-RAA-pion-2760.eps, width=0.45\textwidth, clip=}}
 \subfigure{\label{fig:plot-RAA-jet-2760}
  \epsfig{file=plot-RAA-jet-2760.eps, width=0.45\textwidth, clip=}}
  \caption{(Color online) The nuclear modification factor $R_\mathrm{AA}$ of single hadron at (a) RHIC~\cite{Adare:2012wg} and (b) the LHC~\cite{CMS:2012aa,Abelev:2012hxa}, and single inclusive jet at the LHC~\cite{Khachatryan:2016jfl}.}
  \label{fig:plot-RAA}
\end{figure}

\begin{figure}[tb]
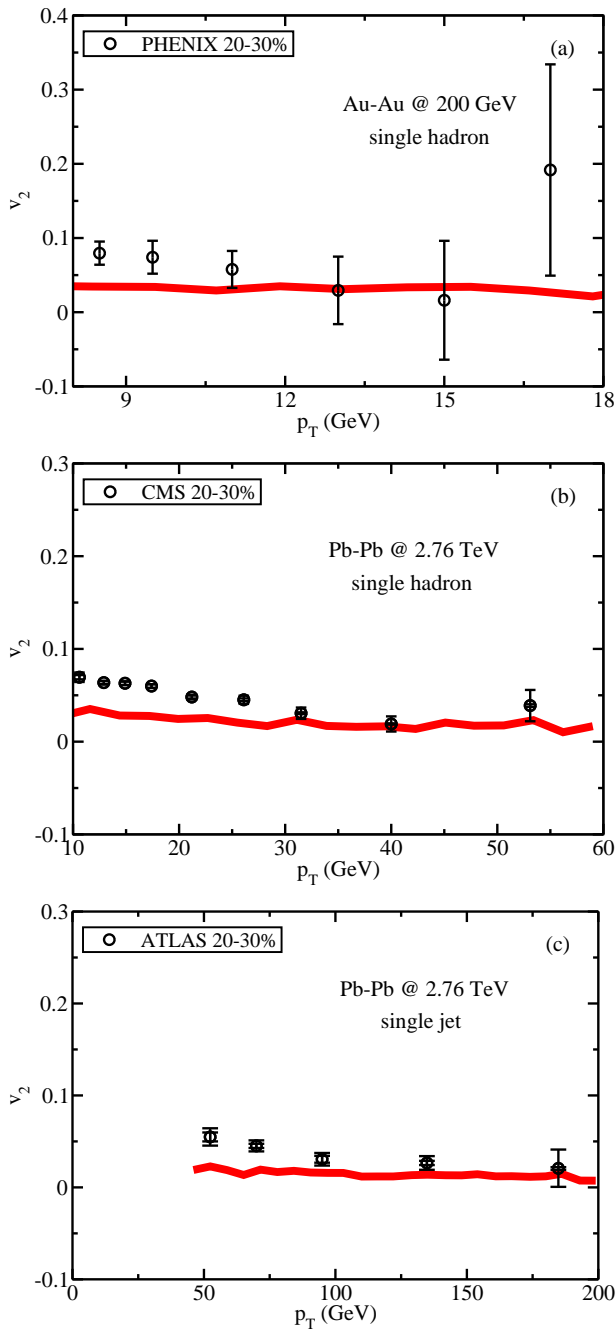

 \subfigure{\label{fig:plot-v2-pion-200}
  \epsfig{file=plot-v2-pion-200.eps, width=0.45\textwidth, clip=}}
 \subfigure{\label{fig:plot-v2-pion-2760}
  \epsfig{file=plot-v2-pion-2760.eps, width=0.45\textwidth, clip=}}
 \subfigure{\label{fig:plot-v2-jet-2760}
  \epsfig{file=plot-v2-jet-2760.eps, width=0.45\textwidth, clip=}}
  \caption{(Color online) The elliptic flow coefficient $v_2$ of single hadron at (a) RHIC~\cite{Adare:2010sp} and (b) the LHC~\cite{Chatrchyan:2012xq}, and single inclusive jet at the LHC~\cite{Aad:2013sla}.}
  \label{fig:plot-v2}
\end{figure}

With this setup, we first present the $p_\mathrm{T}$ spectra of single hadrons and jets from \textsc{Matter} vacuum showers in Fig.~\ref{fig:plot-pp_base}. They are consistent with experimental data at both RHIC and the LHC, and therefore serve as reliable baselines of proton-proton collisions for our study of medium modification in nucleus-nucleus collisions. The nuclear modification factor $R_\mathrm{AA}$ is utilized to quantify the medium modification of hadron and jet production, and the elliptic flow coefficient $v_2$ is used to quantify the azimuthal anisotropy of the medium modification in the event plane. They are calculated as follows:
\begin{align}
& R_\mathrm{AA}(p_\mathrm{T})\equiv\frac{1}{N_\mathrm{coll}}\frac{{dN^\mathrm{AA}}/{dp_\mathrm{T}}}{{dN^\mathrm{pp}}/{dp_\mathrm{T}}}, \\
& v_2(p_\mathrm{T})\equiv\langle \cos(2\phi)\rangle=\left\langle\frac{p_x^2-p_y^2}{p_x^2+p_y^2}\right\rangle,
\end{align}
where $\langle ... \rangle$ denotes averaging over all hadrons or jets inside the same collision system.

In Fig.~\ref{fig:plot-RAA}, we present our numerical results of single hadron and jet $R_\mathrm{AA}$ in 200~AGeV Au-Au collisions at RHIC and 2.76~ATeV Pb-Pb collisions at the LHC. Two centrality regions are included for each observable. As shown in Fig.~\ref{fig:plot-RAA}, \textsc{Matter} calculation provides reasonable descriptions of both single hadron and jet $R_\mathrm{AA}$ across different centralities from RHIC to the LHC. The $\hat{q}_0$ values (at $s_0=96~\mathrm{fm}^{-3}$) of the gluon jet transport coefficient we use are 1.8~(3.0)~GeV$^2$/fm at RHIC and 0.9~(1.5)~GeV$^2$/fm at the LHC when the GW (AZZ) medium-induced splitting function is applied. These are equivalent to the quark transport coefficient $\hat{q}_q/T^3$ around 3.1~(5.2) at RHIC and 1.6~(2.6) at the LHC. As discussed below Eq.~(\ref{eq:medP}), to provide the same numerical result, different values of $\hat{q}$ are required when using GW and AZZ splitting functions.

A closer investigation of Fig.~\ref{fig:plot-RAA-pion-2760} indicates our calculation may slightly overestimate the single hadron $R_\mathrm{AA}$ at low $p_\mathrm{T}$ (below 20~GeV) because the parton showers at low virtuality scale (below $t_\mathrm{min}$) are not included in \textsc{Matter}. This may underestimate the energy loss of partons at low $p_\mathrm{T}$. Improvement is expected when the multi-stage approach of jet evolution is implemented, as developed in Ref.~\cite{Cao:2017zih}. On the other hand, such discrepancy at low $p_\mathrm{T}$ is not manifest in Fig.~\ref{fig:plot-RAA-pion-200} for the single hadron $R_\mathrm{AA}$ at RHIC and in Fig.~\ref{fig:plot-v2-jet-2760} for the single jet $R_\mathrm{AA}$ at LHC. This results from the relatively softer (more rapidly decreasing) hadron spectra at RHIC and jet spectra at the LHC (as shown in Fig.~\ref{fig:plot-pp_base} compared to the hadron spectra at the LHC) that lead to their flatter shapes of $R_\mathrm{AA}$ within the $p_\mathrm{T}$ regimes under investigation. Note that for the medium modified splitting function Eq.~(\ref{eq:totP}) to be reliable, the medium induced virtuality scale $\hat{q}L$ is expected to be a perturbative correction to the parton virtuality scale $t$. Since the average temperature $T$ (thus $\hat{q}$), as well as the medium length $L$ at the LHC is larger than that at RHIC, even higher parton energy (or $p_\mathrm{T}$) is required at the LHC than is required at RHIC for this parton shower formalism to be valid.

The elliptic flow coefficient $v_2$ of both single hadrons and jets are presented in Fig.~\ref{fig:plot-v2}. While our results are consistent with experimental data at high $p_\mathrm{T}$, deviation is observed as $p_\mathrm{T}$ decreases, because of the lack of parton-medium interaction at low virtuality scale. In addition, non-trivial temperature dependence of the jet transport coefficient other than being proportional to $s$ (or $T^3$) may also affect the anisotropy of jet energy loss through the QGP~\cite{Xu:2014tda,Das:2015ana,Cao:2016gvr,Zigic:2018smz,Djordjevic:2019tdu}.

\section{Summary and outlook}
\label{sec:summary}

In this work, we have presented a virtuality-ordered parton shower model -- \textsc{Matter} -- for high energy nuclear collisions. Parton masses (or virtualities) are sampled based on the Sudakov form factor that is constructed by a combination of vacuum and medium-induced splitting functions. The medium-induced part is treated as a perturbation on the vacuum part in nucleus-nucleus collisions and is adopted from the higher-twist energy loss formalism. The full parton splitting process is simulated by determining the longitudinal momentum fractions of the daughter partons taken from the parent via the splitting function, and the transverse momenta of the daughters with respect to the parent via their mass difference. The spacetime structure of parton showers is built into the formalism by sampling the propagation length of each splitting from a Gaussian distribution with an expectation value as the average lifetime of the virtual parton. With this setup, each parton starts with a maximum possible virtuality as its initial energy and evolves through \textsc{Matter} until all daughter partons approach the lower virtuality limit $t_\mathrm{min}=1~\mathrm{GeV}^2$. With the implementation of the vacuum splitting function alone, \textsc{Matter} provides consistent results with \textsc{Jetset} for the daughter parton spectra evolving from single quarks. The $p_\mathrm{T}$ spectra of both single hadrons and jets from \textsc{Matter} vacuum showers are also consistent with experimental data from proton-proton collisions at RHIC and the LHC.

To investigate the nuclear modification of hadron and jet production in heavy-ion collisions, we couple \textsc{Matter} to a hydrodynamic medium that provides the realistic evolution profile of local entropy density and flow velocity of the QGP. The local jet transport coefficient, that enters the path integral of the medium-induced splitting function, is assumed to be proportional to the entropy density in the local rest frame of the fluid cell through which the given parton propagates, and then converted into the global computational frame according to the boost from the fluid flow. Within this framework, our calculation provides reasonable descriptions of the nuclear modification factor $R_\mathrm{AA}$ of both high $p_\mathrm{T}$ single hadrons and jets, with gluon transport coefficients set as $\hat{q}_0=1.8~(3.0)~\mathrm{GeV}^2/\mathrm{fm}$ at RHIC and $0.9~(1.5)~\mathrm{GeV}^2/\mathrm{fm}$ at the LHC (scaled at $s_0=96~\mathrm{fm}^{-3}$) when the GW (AZZ) medium-induced splitting function is applied. The elliptic flow coefficient $v_2$ of both single hadrons and jets agree with experimental data at high $p_\mathrm{T}$ but tends to be underestimated as $p_\mathrm{T}$ decreases. This may be partly due to the lack of parton dynamics at low virtuality within the \textsc{Matter} framework, and partly due to our minimal assumption of the entropy density dependence of the jet transport coefficient.

This work contributes to a more quantitative understanding of the nuclear modification of jet production in relativistic heavy-ion collisions. It extends semi-analytically solving the DGLAP evolution equation so that non-collinear splittings of finite energy partons and their effects on the daughter parton trajectories through the medium can be conveniently incorporated for jet showers. It also extends the medium modification of energetic partons to high virtuality scale, a region of the phase space that is usually neglected by event generators based on on-shell transport models. On the other hand, as discussed earlier, parton showers at low virtuality scale is beyond the description of this Sudakov-based formalism, and thus applying \textsc{Matter} alone is not sufficient either to describe jet observables especially at low $p_\mathrm{T}$. A complete description of jet evolution should involve dynamics at various scales. A multi-stage approach has been introduced by the JETSCAPE collaboration work~\cite{Cao:2017zih} where highly virtual partons first evolve in \textsc{Matter} in virtuality order and then smoothly transit to transport models, such as \textsc{Lbt} and \textsc{Martini}, at low virtuality. This approach has been successfully implemented in a static thermal medium (or a brick) and is ready to be extended to realistic heavy-ion collisions. In addition, jet observables discussed in this work are constructed at the parton level. Apart from the fragmentation mechanism for hard hadron production, the coalescence mechanism~\cite{Han:2016uhh} will be included as well in our upcoming work for soft hadrons and realize study of jets at the hadron level.

\section*{Acknowledgments}

We are grateful to discussions with G.-Y. Qin and M. Kordell, and computing resources provided by Duke University and the Open Science Grid (OSG). This work is funded in part by the U.S. Department of Energy (DOE) under grant number DE-SC0013460 and in part by the National Science Foundation (NSF) under grant number ACI-1550300 within the framework of the JETSCAPE collaboration.

\bibliographystyle{h-physrev5}
\bibliography{SCrefs}

\end{document}